\documentclass[3p,times,twocolumn]{elsarticle}

\usepackage{ecrc}

\usepackage{hyperref} 
\usepackage{amsmath} 

\volume{260}

\firstpage{130}

\journalname{Nuclear and Particle Physics Proceedings}

\runauth{Zhiqing Zhang}

\jid{nuphbp}

\jnltitlelogo{Nuclear Physics B Proceedings Supplement}

\usepackage{amssymb}

\biboptions{sort&compress}

\usepackage[figuresright]{rotating}

\begin{document}

\begin{frontmatter}


\title{Tau Decays and $\alpha_s$}
\author{Zhiqing Zhang\cortext[a]{{\it Email address: }{\tt zhangzq@lal.in2p3.fr} (Zhiqing Zhang)}}
\address{Laboratoire de l'Acc\'el\'erateur Lin\'eaire, Univ.\ Paris-Sud 11 et IN2P3/CNRS, France}

\dochead{}





\begin{abstract}
The evolution of the determination of the strong coupling constant $\alpha_s$ from the leptonic branching ratios, the lifetime, and the invariant mass distributions of the hadronic final state of the $\tau$ lepton over the last two decades is briefly reviewed. The improvements in the latest ALEPH update are described in some detail. Currently this is one of the most precise $\alpha_s$ determinations. Together with the other determination at the $Z$ boson mass pole, they constitutes the most accurate test of the asymptotic freedom in QCD.
\end{abstract}


\end{frontmatter}


\section{Introduction}
\label{intro}

The strong coupling constant $\alpha_s$ is one of the three fundamental gauge coupling constants of the Standard Model of particle physics and it is thus of utmost importance to measure it with high precision. 
Hadronic $\tau$ decays provide an especially clean environment for the study of QCD effects
and in particular the determination of $\alpha_s$.
At the $\tau$ mass scale, its value is predicted to be larger than that at higher energy due to the energy dependence based on the renormalization group equations. Consequently, the hadronic branching ratio of the tau decays $B(\tau^-\to h^-\nu_\tau)$ is expected to be much larger than the predicted value of 3/5 at the quark-parton level and ignoring quark masses. The sizable deviation is indeed measured and this has been used together with the measured spectral functions of the $\tau$ hadronic decays to determine $\alpha_s(m_\tau^2)$ as well as small non-perturbative contributions using a method proposed in~\cite{plb289}.  

Measurements performed by ALEPH~\cite{aleph93,aleph98,aleph05,aleph13}, CLEO~\cite{cleo95} and OPAL~\cite{opal99} are briefly reviewed in the following together with other determinations, 
focusing on the latest ALEPH results. Over the years, the two main improvements in the measurement of $\alpha_s(m^2_\tau)$ were from  the increased precision of the branching ratio and spectral function measurements and the higher-order perturbative calculation.

\section{Experimental inputs and theoretical basis}
\label{inputs}

Experimentally, one measures branching ratios of all leptonic, non-strange and strange hadronic decay modes of the $\tau$ decays (see e.g.~\cite{aleph05}). The non-strange decay modes can be further decomposed into vector and axial-vector components, corresponding to channels with an even and odd number of neutral or charged pions, respectively. The ratio of the total hadronic tau decay width to the leptonic one, $R_\tau=\Gamma(\tau^-\to h^- \nu_\tau)/\Gamma(\tau^-\to e^-\bar{\nu}_e\nu_\tau)=R_{\tau, V}+R_{\tau, A}+R_{\tau, S}=(1-B_e-B_\mu)/B_e=1/B_e^{\rm uni}-1.9726$, can thus be determined with a precise universal electronic branching fraction, $B_e^{\rm uni}$, only. It is an average over the directly measured $B_e$ and two derived ones from the measured $B_\mu$ and tau lifetime $\tau_\tau$.

In addition to the measured branching ratios, spectral functions of the V, A and S components are also measured. These are invariant mass spectra normalized to their branching ratios and corrected for $\tau$ decay kinematics, e.g. for V and A,
\begin{equation*} 
v_1(s)/a_1(s)\propto 
\frac{B_{V/A}}{B_e}\frac{dN_{V/A}}{N_{V/A}ds}
\frac{m^2_\tau}{\left(1-\frac{s}{m^2_\tau}\right)^2\left(1+\frac{2s}{m^2_\tau}\right)}\,.
\end{equation*}

On the theoretical side, using the Operator Product Expansion (OPE)~\cite{ope}, the inclusive vector and axial-vector $\tau$ hadronic width and spectral functions can be expressed as a function of $\alpha_s$ and non-perturbative phenomenological operators~\cite{npb373}: $R_{\tau, V/A}(s_0)\propto |V_{ud}|^2 S_{\rm EW} \left(1+\delta^{(0)}+\delta^\prime_{\rm EW}+\delta_{V/A}^{(2,m_q)}+\sum_{D=4,6,8,\cdots}\delta_{V/A}^{(D)}\right)$, where $s_0=m^2_\tau$, $V_{ud}$ denotes the CKM weak mixing matrix element, $S_{\rm EW}$ accounts for electroweak radiative corrections~\cite{prl61}, the term $\delta^{(0)}$ is the dominant massless perturbative contribution, $\delta^\prime_{\rm EW}=0.0010$ corresponds to the residual non-logarithmic electroweak correction~\cite{prd42}, $\delta^{(2,m_q)}$ arises from quark masses which is smaller than 0.1\% for the $u,d$ quarks, and $\delta^{(D)}$ are the OPE terms in powers of $s^{D/2}_0$. Depending on how the perturbative series is truncated, there are different schemes in the $\alpha_s$ expansion. The two most used schemes are the Fixed-Order Perturbation Theory (FOPT) expansion and the Contour-Improved Perturbation Theory (CIPT) expansion~\cite{plb286}. 

The parameter of interest is $\alpha_s(m^2_\tau)$, if one only considers 3 non-perturbative terms $\delta^{(D)}$ with $D=4,6,8$, one has 4 free parameters to determine. This is achieved by a simultaneous fit to $R_{\tau, V/A}$ and 4 spectral moments $R_{\tau, V/A}^{kl}=\int_0^{m^2_\tau}ds\left(1-\frac{s}{m^2_\tau}\right)^k\left(\frac{s}{m^2_\tau}\right)^l\frac{dR_{\tau, V/A}}{ds}$ with $k=1$ and $l=0,\cdots,3$. Only 4 spectra moments are used because of the strong correlation. In practice the normalized spectral moments $D_{\tau, V/A}^{kl}\equiv R_{\tau, V/A}^{kl}/R_{\tau, V/A}$ are used in order decouple the normalization from the shape of the $\tau$ spectral functions.

\section{Measurements of $\alpha_s(m^2_\tau)$}
\label{measure}

The measurement of $\alpha_s$ at the tau mass scale was carried out for the first time by ALEPH in 1993 using about 8500 $\tau$ decays~\cite{aleph93}. This is followed by CLEO in 1995~\cite{cleo95} and by OPAL in 1999~\cite{opal99} using the same technique.

ALEPH has also made several updates. The update in 1998~\cite{aleph98} was based on about 30 times of more data than those used in~\cite{aleph93} and $\alpha_s$ was determined for the first time with three independent spectral functions $V$, $A$ and $V+A$. In addition, the $V-A$ spectral function also allowed the evaluation of the finite energy chiral sum rules up to the $\tau$ mass. The approach of the OPE was also tested experimentally studying the evolution of the $\tau$ hadronic widths to masses smaller than the $\tau$ mass. The update in 2005~\cite{aleph05} used the full LEP-1 data set with the final branching ratio measurements based on extensive systematic studies. 

Using the same data, two additional improvements were realized. The first one in 2008~\cite{ddhmz08} includes the new fourth-order perturbative coefficient~\cite{prl101} which reduces the dominant theoretical uncertainty due to the truncation of the series by 20\% with respect to earlier determinations based on the third-order expansion, better separation between vector and axial-vector spectral functions by using the new precise $e^+e^-$ annihilation cross section measurements from Babar, and improved results from Babar and Belle on $\tau$ branching ratios involving kaons. The second one in 2013~\cite{aleph13} updated the non-strange spectral functions by using a new and more robust method~\cite{unfold} to unfold the measured mass spectra from detector effects. The procedure also corrected a problem in the correlations between the unfolded mass bins missing in the previous spectral functions~\cite{boito}. 

\begin{figure}[htb]
\begin{center}
\includegraphics[width=.45\textwidth]{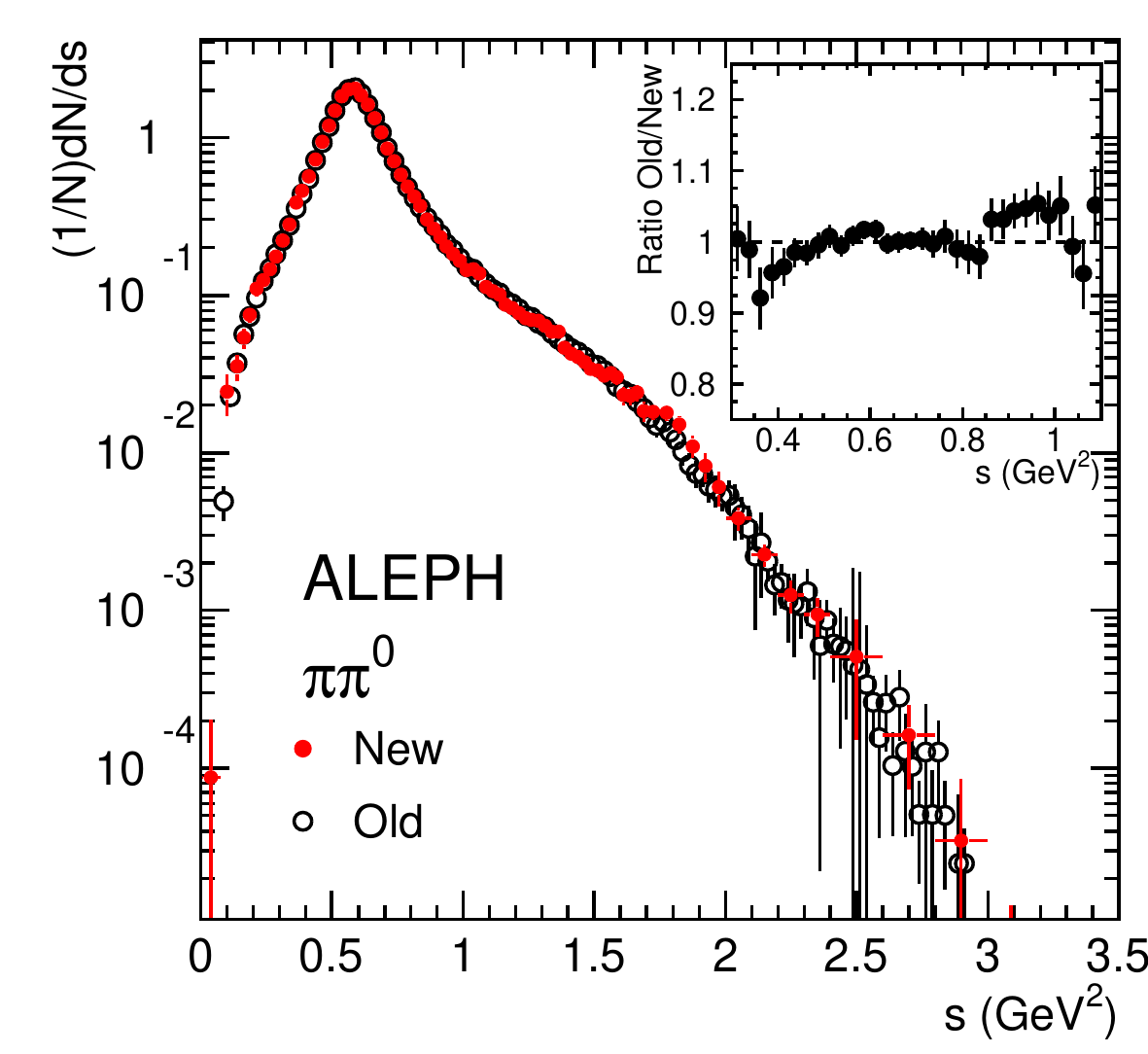}
\includegraphics[width=.45\textwidth]{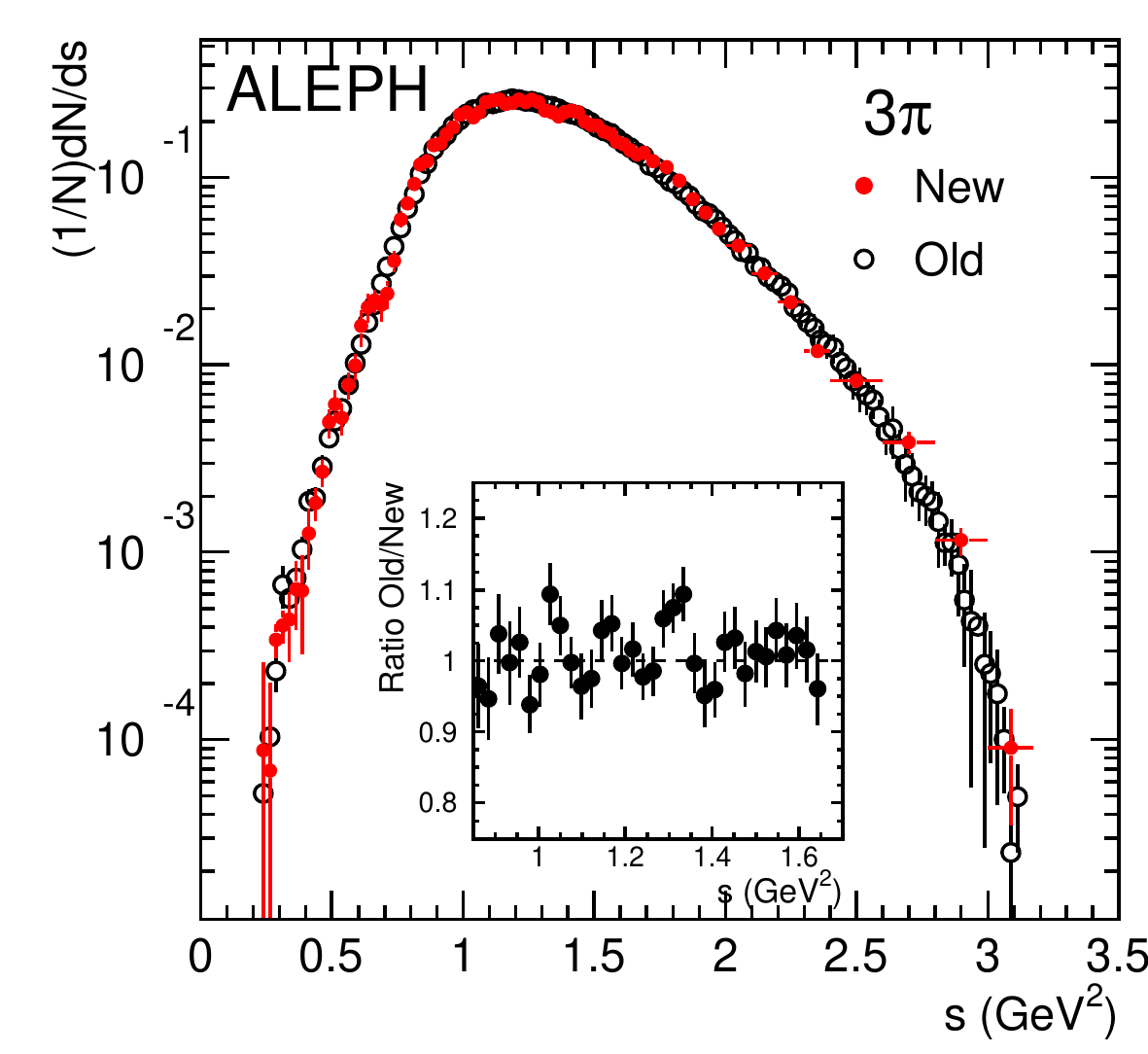}
\end{center}\vspace{-7mm}\caption{Comparion of the new unfolded spectral functions (full circles) with those obtained in~\cite{aleph05} (open circles). The error bars shown include statistical and all systematic uncertainties. The error bars in the insets, showing the old-to-new ratios, are those of the newly unfolded spectra.} 
\label{fig:sf_comp}
\end{figure}
The comparison of the new spectral functions with the old one is shown in Fig.~\ref{fig:sf_comp} for two example cases. 
Reasonable agreement is found everywhere except for differences at the few percent level in the $\pi\pi^0$ mode near threshold and in the $0.8-1.0\,{\rm GeV}^2$ region.

\begin{figure}[htbp]
\begin{center}
\includegraphics[width=.43\textwidth]{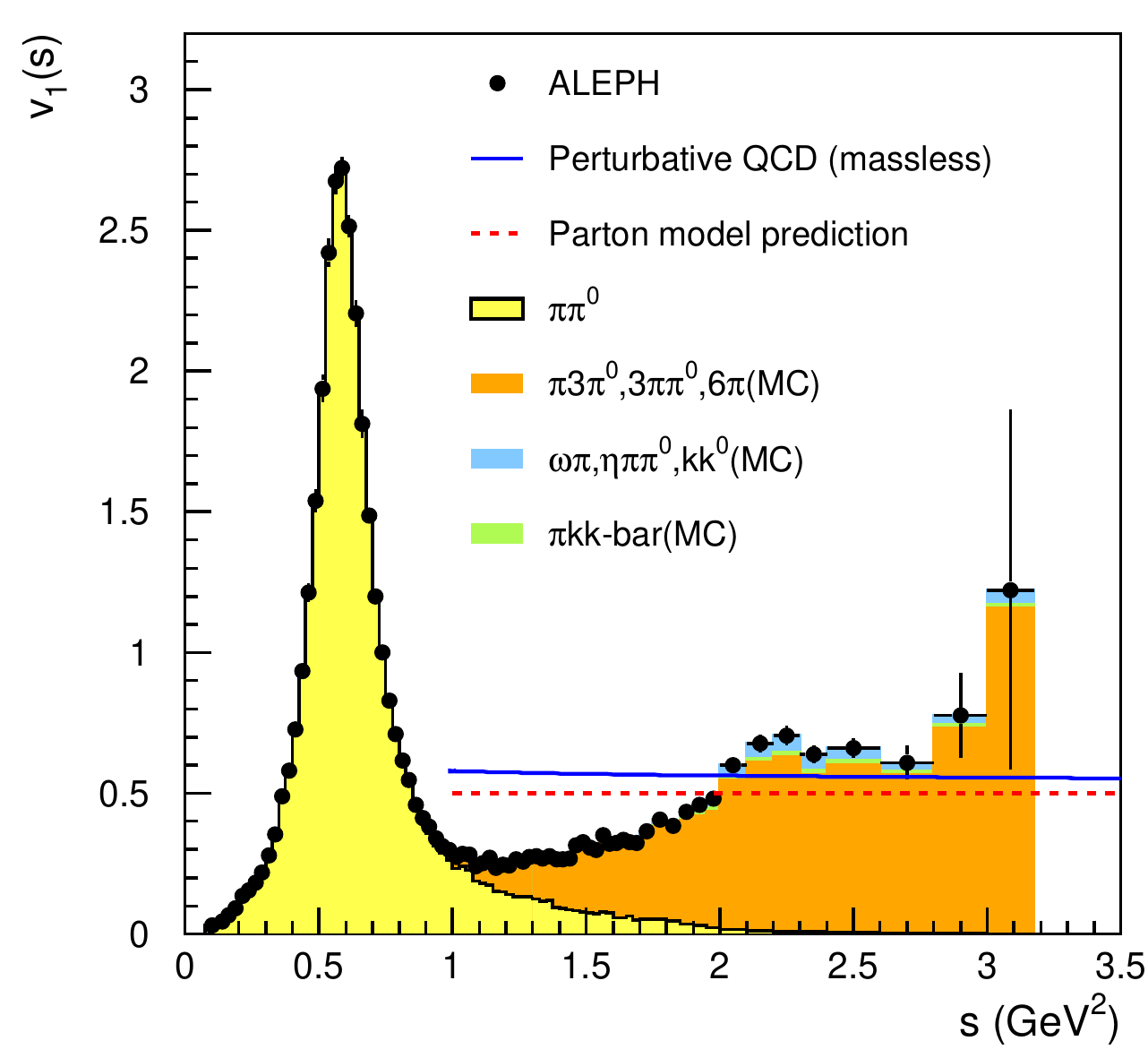}
\includegraphics[width=.43\textwidth]{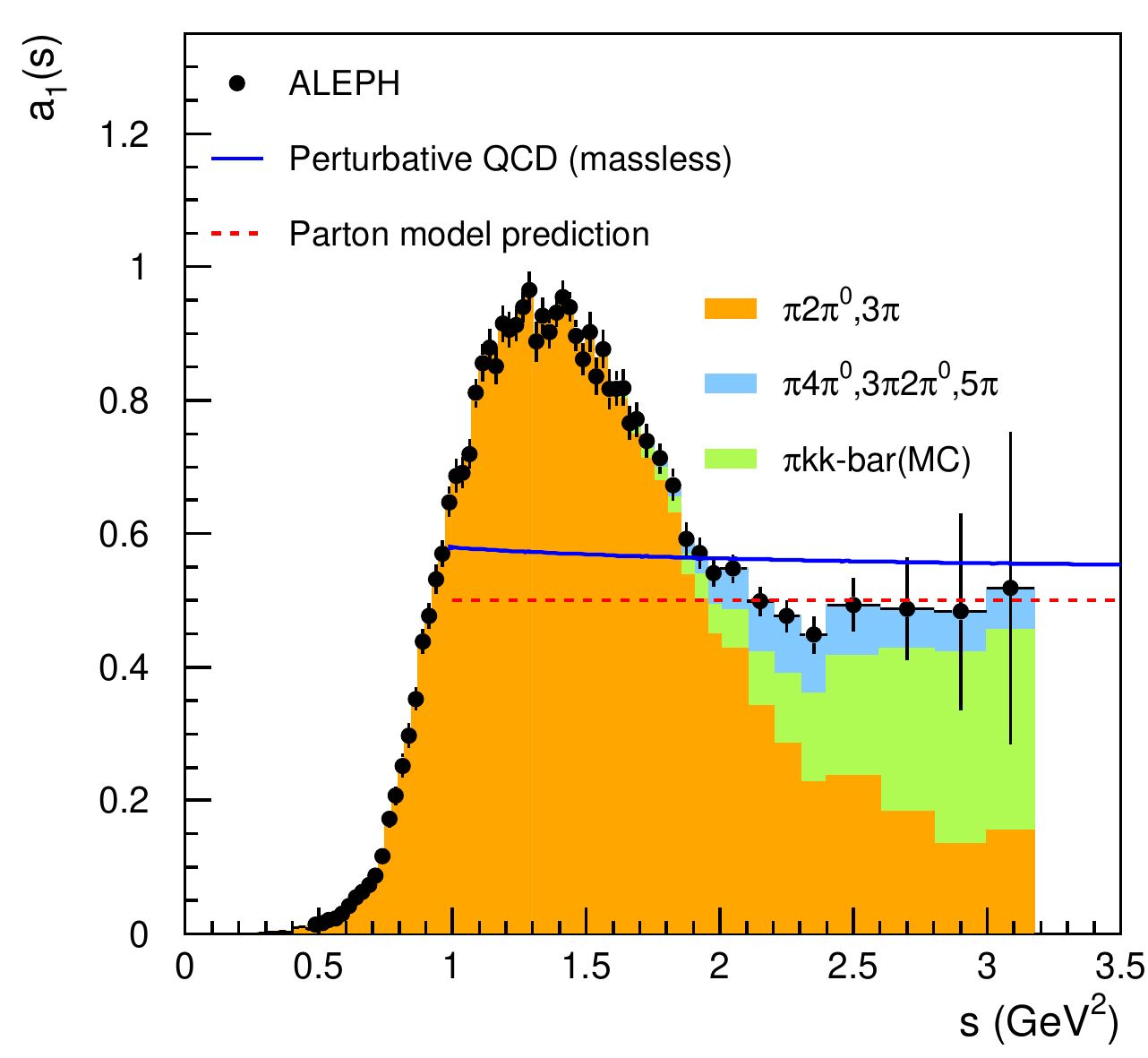}
\includegraphics[width=.43\textwidth]{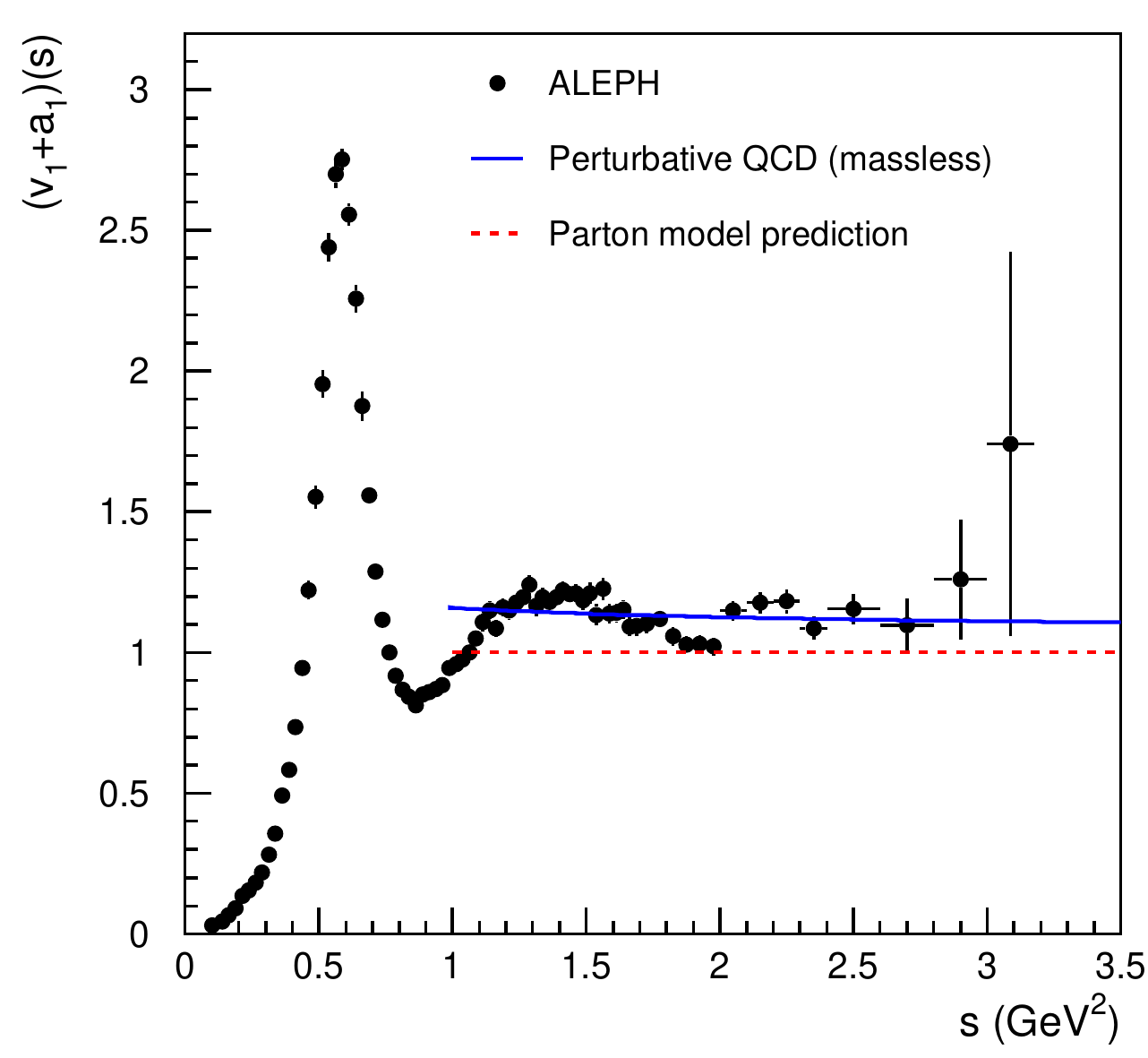}
\end{center}\vspace{-7mm}\caption{Updated ALEPH $\tau$ spectral functions. The shaded areas indicate the contributions from the exclusive $\tau$ decay channels, where the shapes of the contributions labelled ``MC" are taken from the MC simulation. The lines show the predictions from the naive parton model and from massless perturbative QCD using $\alpha_s(M^2_Z)=0.120$, respectively.}
\label{fig:newsf}
\end{figure}
The new spectral functions $V$, $A$ and $V+A$ are displayed in Fig.~\ref{fig:newsf}. The dashed line depicts the naive parton model prediction while the solid line shows the massless perturbative QCD prediction using $\alpha_s(M^2_Z)=0.120$. Although the statistical power of the data is weak near the kinematic limit, the trend of the spectral functions clearly indicates the asymptotic region is not reached for $V$ and $A$. On the other hand, the oscillating behavior of the $V+A$ does approximately reach the asymptotic limit. This implies that the non-perturbative effects in $V+A$ are smaller than those from $V$ or $A$ as it was observed indeed from the individual fits to these spectral functions. Consequently, the $\alpha_s(m^2_\tau)$ determination from $V+A$ is believed to be the most robust one among the three spectral functions.

The corresponding CIPT fit result~\cite{aleph13} is $\alpha_s(m^2_\tau)=0.341\pm 0.005_{\rm exp}\pm 0.006_{\rm theo}$, where the first error is experimental and the second the theoretical one. When averaged with the FOPT fit result, one gets  $\alpha_s(m^2_\tau)=0.332\pm 0.005_{\rm exp}\pm 0.011_{\rm theo}$, the larger theoretical error includes half of the CIPT and FOPT difference ($\pm 0.009$), which after evolution to $M^2_Z$ gives $\alpha_s(M^2_Z)=0.1199\pm 0.006_{\rm exp}\pm 0.0012_{\rm theo}\pm 0.0005_{\rm evol}=0.1199\pm 0.0015_{\rm total}$, where the third error is due to the evolution. The result is in excellent agreement with $\alpha_s(M^2_Z)=0.1196\pm 0.0030$, the theoretically most robust precision determination (also based on fourth-order computation) from the global fit (Gfitter) to electroweak data at the $Z$ mass scale~\cite{gfitter}. 

\begin{figure}[htb]
\begin{center}
\includegraphics[width=.475\textwidth]{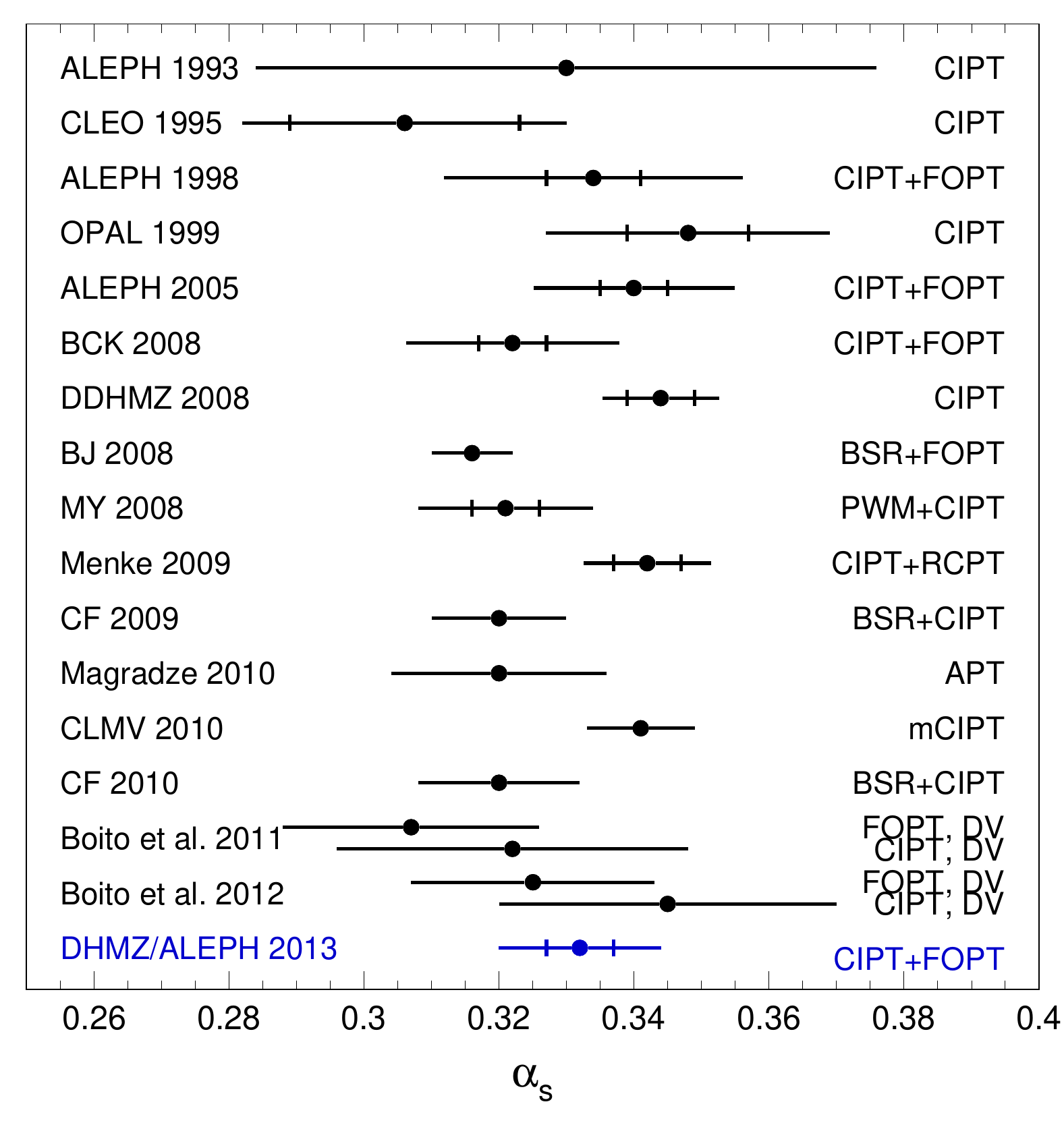}
\end{center}\vspace{-7mm}\caption{Compilation of the determinations of $\alpha_s(m^2_\tau)$ based on the $\tau$ hadronic decays, where CIPT stands for Contour-Improved Perturbation Theory, FOPT for Fixed-Order Perturbation Theory, BSR for Borel Summation of Renormalon series, PWM for Pinched-Weight Moments, RCPT for Renormalon Chain Perturbation Theory, APT for Analytic Perturbation Theory, mCIPT for modified CIPT, DV for Duality Violation. The experimental uncertainty is shown as the inner error bar when available.}
\label{fig3}
\end{figure}

A compilation of these described $\alpha_s(m^2_\tau)$ measurements with other determinations is shown in Fig.~\ref{fig3}. Most of these other determinations~\cite{prl101,bj08,my08,menke09,cf09,magradze10,clmv10,cf10} were based directly or indirectly on the earlier versions of the spectral functions from ALEPH. Only two recent determinations~\cite{boito11,boito12} were obtained using the spectral functions of OPAL. One major difference of the latter determinations with the others is the introduction of duality violation with a model containing four additional parameters. This together with the limited precision of the OPAL data explained larger uncertainties on these $\alpha_s$ determinations.~\footnote{During the workshop, another determination based on the same method was reported using the new spectral functions of ALEPH~\cite{aleph13}, details of the analysis are available after the workshop in~\cite{boito14}. In this article, it was claimed that the original method proposed in~\cite{plb289}, neglecting the nonperturbative contribution at orders higher than $D=8$ and the duality violation effects, could no longer be used given the precision of the data. 
However, the existing studies do not allow to draw an unambiguous conclusion on this point.}

\section{Conclusion}
\label{summary}

The determination of the strong coupling constant $\alpha_s(m^2_\tau)$ at the $\tau$ mass scale using the branching ratios and spectral functions of the non-strange $\tau$ hadronic decays has been briefly described and reviewed. The impact of the revised spectral functions of ALEPH on $\alpha_s(m^2_\tau)$ is found to be small. It is one of the few $\alpha_s$ measurements performed with perturbative predictions known at fourth-order. Because the $\alpha_s$ uncertainty scales with $\alpha_s^2$, when it is extrapolated to the $Z$ mass, it yields one of the most precise $\alpha_s$ determinations. The good agreement with the other independent measurements at the $Z$ mass pole also tests the asymptotic freedom of the QCD with unprecedented precision.

\vspace{5mm}
{\small I am grateful to the fruitful collaboration with my colleagues and friends in particular Michel Davier, Andreas Hoecker and Bogdan Malaescu.}





\bibliographystyle{elsarticle-num}



\end{document}